\documentclass[aps,prb,reprint,superscriptaddress,longbibliography,amsmath,amssymb,floatfix]{revtex4-2}

\usepackage{xcolor}
\usepackage{graphicx}  
\usepackage{dcolumn}    
\usepackage{bm}         
\usepackage{hyperref}   
\usepackage[normalem]{ulem}
\usepackage{pdfcomment}

\graphicspath{{./}}   

\newcommand{\figStructure}{fig1.png}        
\newcommand{\figOrbital}{fig2.png}          
\newcommand{\figSpinsplit}{fig3.png}        
\newcommand{\figStype}{fig4.png}        
\newcommand{\figSchematic}{fig5.png}        
\newcommand{\figEnlarged}{fig6.png}         
\newcommand{\figSplitting}{fig7.png}        
\newcommand{\figAnticrossing}{fig8.png}     
\newcommand{\figChartable}{fig9.png}        

\begin{document}

\title{First-principles calculations of spin-split bands in chiral hybrid organic--inorganic perovskites (\textit{R}/\textit{S}-PEA)PbI$_3$ and (\textit{R}/\textit{S}-NEA)PbI$_3$}

\author{Tetsuya Furukawa}
\email{tetsuya.furukawa.c1@tohoku.ac.jp}
\affiliation{Institute for Materials Research, Tohoku University, Sendai 980-8577, Japan}

\author{Kazushi Nakano}
\affiliation{Department of Applied Physics, Tokyo University of Science, Tokyo 125-8585, Japan}

\author{Youta Suzuki}
\affiliation{Department of Applied Physics, Tokyo University of Science, Tokyo 125-8585, Japan}

\author{Takumi Kaneko}
\affiliation{Department of Applied Physics, Tokyo University of Science, Tokyo 125-8585, Japan}

\author{Ayumi Ishii}
\affiliation{School of Advanced Science and Engineering, Waseda University, Tokyo 169-8555, Japan}

\author{Tetsuaki Itou}
\email{tetsuaki.itou@rs.tus.ac.jp}
\affiliation{Department of Applied Physics, Tokyo University of Science, Tokyo 125-8585, Japan}

\date{\today}

\begin{abstract}
Chiral hybrid organic--inorganic perovskites provide a promising platform for investigating the physics of chirality-driven spin-split bands because they combine robust molecular chirality with strong spin--orbit coupling from heavy inorganic ions.
First-principles calculations including spin--orbit coupling are performed for the one-dimensional chiral perovskites (\textit{R}/\textit{S}-PEA)PbI$_3$ and (\textit{R}/\textit{S}-NEA)PbI$_3$ to compare their spin-split band structures and to identify the factors controlling their differences.
In (\textit{R}/\textit{S}-PEA)PbI$_3$, the lowest conduction bands predominantly consist of Pb orbitals, whereas in (\textit{R}/\textit{S}-NEA)PbI$_3$, they are formed by hybridization between Pb orbitals and the lowest unoccupied molecular orbital of NEA.
Both compounds exhibit pronounced spin splitting near the valence-band maximum and conduction-band minimum. The effective spin splitting of the edges of the valence bands is stronger in (\textit{R}/\textit{S}-NEA)PbI$_3$, despite similar linear-in-$k$ splitting coefficients near the relevant high-symmetry points.
This enhancement originates from larger gaps induced by spin--orbit coupling at high-symmetry points and band (anti)crossings in the multiband structure.
For a given molecular handedness, the PEA- and NEA-based compounds exhibit opposite spin textures, consistent with the opposite chiral distortions of the [PbI$_6$]$^{4-}$ octahedra and with the previously observed opposite signs of circular dichroism.
Group-theoretical analysis for the nonsymmorphic space group $P2_12_12_1$ further accounts for band sticking, symmetry-enforced degeneracies, and the disappearance of spin polarization at specific Brillouin-zone-boundary points.
These results provide a solid foundation for future studies of chirality-dependent electromagnetic responses, including circular dichroism, in chiral hybrid organic--inorganic perovskites.
\end{abstract}

\maketitle

\section{\label{sec:intro}Introduction}
Chirality is a concept that distinguishes between right- and left-handedness.
A chiral structure is defined as one that cannot be superimposed onto its mirror image by any (proper) rotation~\cite{Barron2020}.
In a chiral crystal, the electron energy is shifted by the electronic chirality term $\Delta E = \beta(\bm{p}\cdot\bm{s})$, where $\bm{p}$ is the linear momentum, $\bm{s}$ is the spin angular momentum, and $\beta$ is the time-reversal-invariant pseudoscalar coefficient whose sign is determined by structural chirality.
The locking between linear momentum and spin angular momentum induces spin-split bands in a chiral crystal.
Through these spin-split bands, chiral crystals give rise to intriguing and useful phenomena, such as optical activity~\cite{Landau1984}, circular dichroism (CD)~\cite{Landau1984}, second-order nonlinear optical effects~\cite{Boyd2003}, and current-induced magnetization (the Edelstein effect)~\cite{Edelstein1990}.

For a material to serve as a good platform for exploring chirality-related electronic phenomena, several conditions are crucial:
(i)~a well-defined structural chirality with a large distinction between enantiomorphic crystals, (ii)~the ability to obtain and selectively synthesize homochiral crystals, and (iii)~strong spin--orbit coupling (SOC) that can transfer structural chirality to the electronic spin degrees of freedom.
Only a limited number of materials satisfy these conditions well, and thus the physics of chirality has advanced primarily through several prototypical systems such as elemental tellurium~\cite{Hirayama2015,Vorobev1979,Furukawa2017,Furukawa2021}.

\begin{figure}[!t]
  \includegraphics[width=\columnwidth,clip]{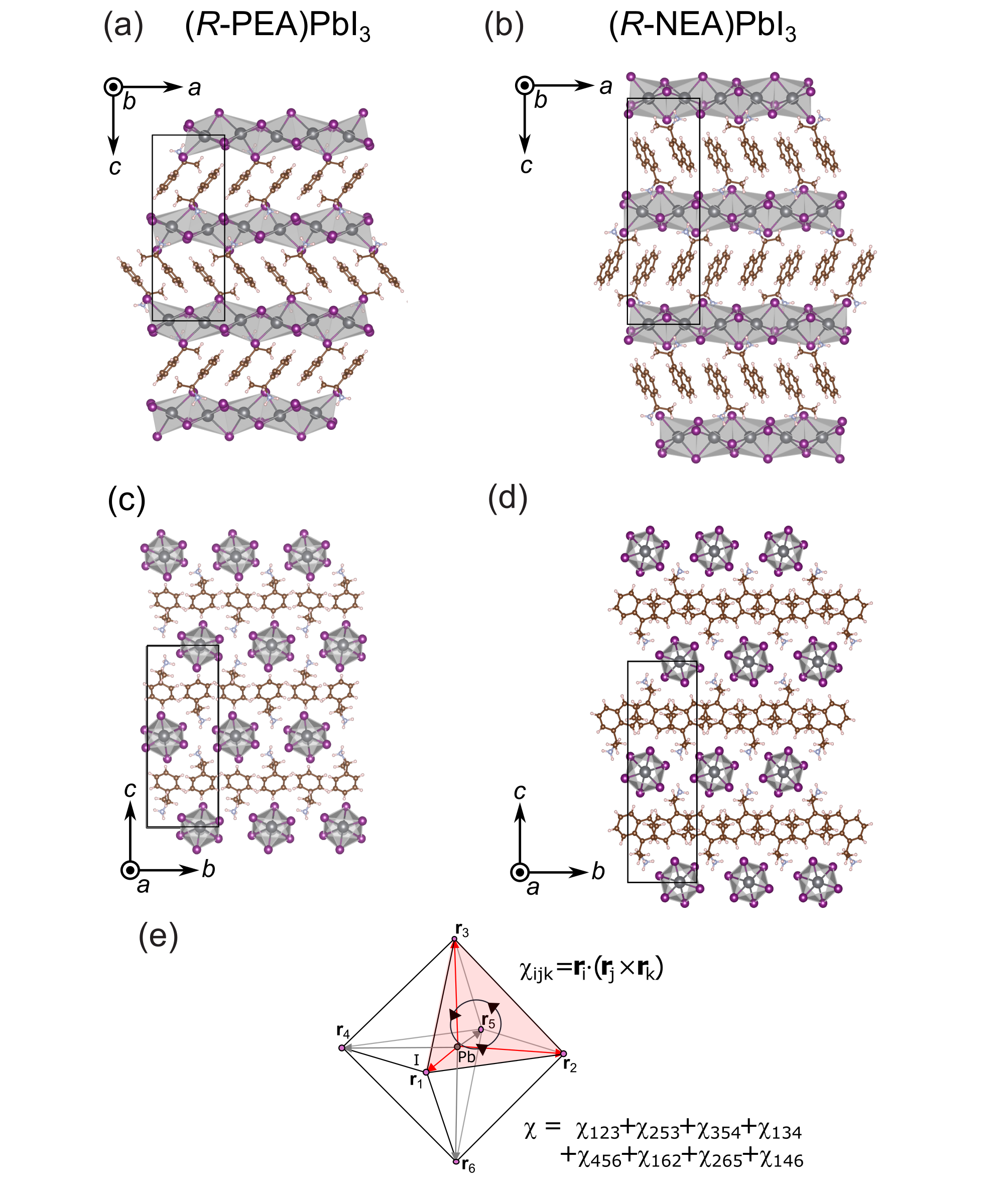} 
  \caption{\label{fig:structure}
    Crystal structures of (\textit{R}-PEA)PbI$_3$ and (\textit{R}-NEA)PbI$_3$.
    (a),(c) The $b$- and $a$-axis views of (\textit{R}-PEA)PbI$_3$, respectively.
    (b),(d) The $b$- and $a$-axis views of (\textit{R}-NEA)PbI$_3$, respectively.
    (e) Schematic definition of the indicator of structural chirality $\chi$ of the [PbI$_6$]$^{4-}$ octahedron.}
\end{figure}

Chiral hybrid organic--inorganic perovskites (HOIPs), composed of chiral organic molecules and heavy inorganic ions, have recently emerged as promising chiral materials~\cite{Long2020, Pietropaolo2022}.
In these systems, the robust structural chirality of the organic molecules can dominate the overall crystal chirality, while the heavy inorganic atoms provide strong SOC that is often lacking in pure organic compounds.
Combining the advantages of both material classes enables exploration of new chirality-related electronic phenomena.
Pb-based chiral HOIPs have attracted particular attention because of their large CD signals, namely, the difference in absorption between right- and left-handed circularly polarized light~\cite{Ahn2017,Ahn2020,Chen2019,Ishii2020}.
Among them, one-dimensional (1D) systems such as (\textit{R}/\textit{S}-PEA)PbI$_3$ and (\textit{R}/\textit{S}-NEA)PbI$_3$ stand out because of their exceptionally large CD responses~\cite{Chen2019,Ishii2020}.
These materials consist of alternating organic and inorganic layers stacked along the crystallographic $c$ axis: the organic layers comprise chiral amine molecules---(\textit{R})-(+)- or (\textit{S})-(-)-1-phenylethylamine (\textit{R}/\textit{S}-PEA, also called \textit{R}/\textit{S}-MBA), or (\textit{R})-(+)- or (\textit{S})-(-)-1-(1-naphthyl)ethylamine (\textit{R}/\textit{S}-NEA)---and the inorganic layers comprise chains of [PbI$_6$]$^{4-}$ octahedra [Fig.~\ref{fig:structure}(a),(b)]~\cite{Ishii2025}.
The [PbI$_6$]$^{4-}$ octahedra share faces to form chains along the $a$ axis, which are aligned antiparallel to each other along the $b$ axis.
The molecular planes of the chiral amines are oriented approximately along the $a$ axis.
The chirality of the organic molecules induces distortions of the inorganic framework via hydrogen bonding, twisting the [PbI$_6$]$^{4-}$ octahedra chains.
The molecular handedness uniquely determines the handedness of the chiral crystal structure.
The transfer of structural chirality to the electronic states, accompanied by spin--momentum locking, is referred to as \textit{chirality transfer}~\cite{Long2018,Jana2020}.

\begin{table}[t]
  \caption{\label{tab:lattice_constants}
    Lattice constants of (\textit{R}/\textit{S}-PEA)PbI$_3$ and (\textit{R}/\textit{S}-NEA)PbI$_3$.}
  \begin{ruledtabular}
    \begin{tabular}{lccc}
      Compound & $a$ (\AA) & $b$ (\AA) & $c$ (\AA) \\
      \hline
      (\textit{R}-PEA)PbI$_3$ &8.074&8.646&20.862\\
      (\textit{S}-PEA)PbI$_3$ &8.069&8.650&20.885\\
      (\textit{R}-NEA)PbI$_3$ &8.057&8.357&25.318\\
      (\textit{S}-NEA)PbI$_3$ &8.050&8.350&25.295\\
      \end{tabular}
  \end{ruledtabular}
\end{table}

The crystal structures of (\textit{R}/\textit{S}-PEA)PbI$_3$ and (\textit{R}/\textit{S}-NEA)PbI$_3$ both belong to the orthorhombic space group $P2_12_12_1$ (No.~19).
Because this crystal structure is chiral but nonpolar, it hosts a radial spin texture, rather than a Rashba-type circular spin texture found in polar systems.
Large CD signals have been observed for both PEA- and NEA-based compounds; however, the CD magnitude in the NEA compound is approximately twice that in the PEA compound~\cite{Ishii2020}.
This difference cannot be attributed solely to film quality or crystallinity but rather originates from intrinsic differences in the chiral crystal structures themselves~\cite{Ishii2020}.
In addition, for a given molecular handedness, either \textit{R} or \textit{S}, the PEA- and NEA-based materials exhibit CD signals of opposite sign.
The relationship between the handedness of the molecules and the sign of the CD signals remains unclear.
Thus, understanding the distinctions between the spin-split band structures of the NEA and PEA systems is crucial for elucidating the mechanism behind the giant CD.

The band structures of 1D lead-based chiral HOIPs have been investigated using first-principles calculations.
For (\textit{R}/\textit{S}-PEA)PbI$_3$, Wei \textit{et al.}\ reported that spin splitting arises due to the strong SOC of lead, and they attributed the magnitude of the splitting to the local electric field associated with the displacement of Pb from the center of the surrounding iodide octahedron~\cite{Wei2021}.
The band structure of (\textit{R}/\textit{S}-NEA)PbI$_3$ has also been calculated by Xiao \textit{et al.}, and analyses based on a chirality-induced spin--orbit interaction model suggest that octahedral distortions associated with structural chirality contribute to the observed CD~\cite{Xiao2024}.
However, no first-principles calculations including SOC have yet been performed to explicitly determine the spin-split band structure of
(\textit{R}/\textit{S}-NEA)PbI$_3$, and a quantitative comparison between the spin-split bands of NEA and PEA compounds has been lacking.
In addition, the influence of the nonsymmorphic nature of the $P2_12_12_1$ space group has not been discussed in previous studies.
Given that many other chiral HOIPs share the same $P2_12_12_1$-type symmetry, a systematic understanding of spin-split band structures in
nonsymmorphic chiral systems is essential.

In this work, we therefore calculated the spin-split band structure of (\textit{R}/\textit{S}-NEA)PbI$_3$ using first-principles methods and quantitatively compared it with that of (\textit{R}/\textit{S}-PEA)PbI$_3$.
Furthermore, we performed a group-theoretical analysis to elucidate how nonsymmorphic symmetry affects spin splitting and band degeneracies.
Our main findings are as follows:
(1)~the lowest conduction bands are predominantly Pb-derived in (\textit{R}/\textit{S}-PEA)PbI$_3$, whereas those in (\textit{R}/\textit{S}-NEA)PbI$_3$ have mixed Pb-orbital and lowest-unoccupied-molecular-orbital (LUMO) character;
(2)~the effective spin splitting near the edges of valence bands is stronger in (\textit{R}/\textit{S}-NEA)PbI$_3$ because of larger spin-orbit-coupling-induced gaps at high-symmetry points and band (anti)crossings in the multiband structure;
(3)~for a given molecular handedness, either \textit{R} or \textit{S}, the reversal of the spin textures between the PEA- and NEA-based compounds is consistent with the opposite chiral distortions of the [PbI$_6$]$^{4-}$ octahedra and with the previously observed opposite signs of CD; and (4)~nonsymmorphic symmetry accounts for band sticking, symmetry-enforced degeneracies, and the disappearance of spin polarization at specific Brillouin-zone-boundary points.

\section{\label{sec:dft}DFT Calculation}
\begin{figure*}[t]
  \includegraphics[width=18cm,clip]{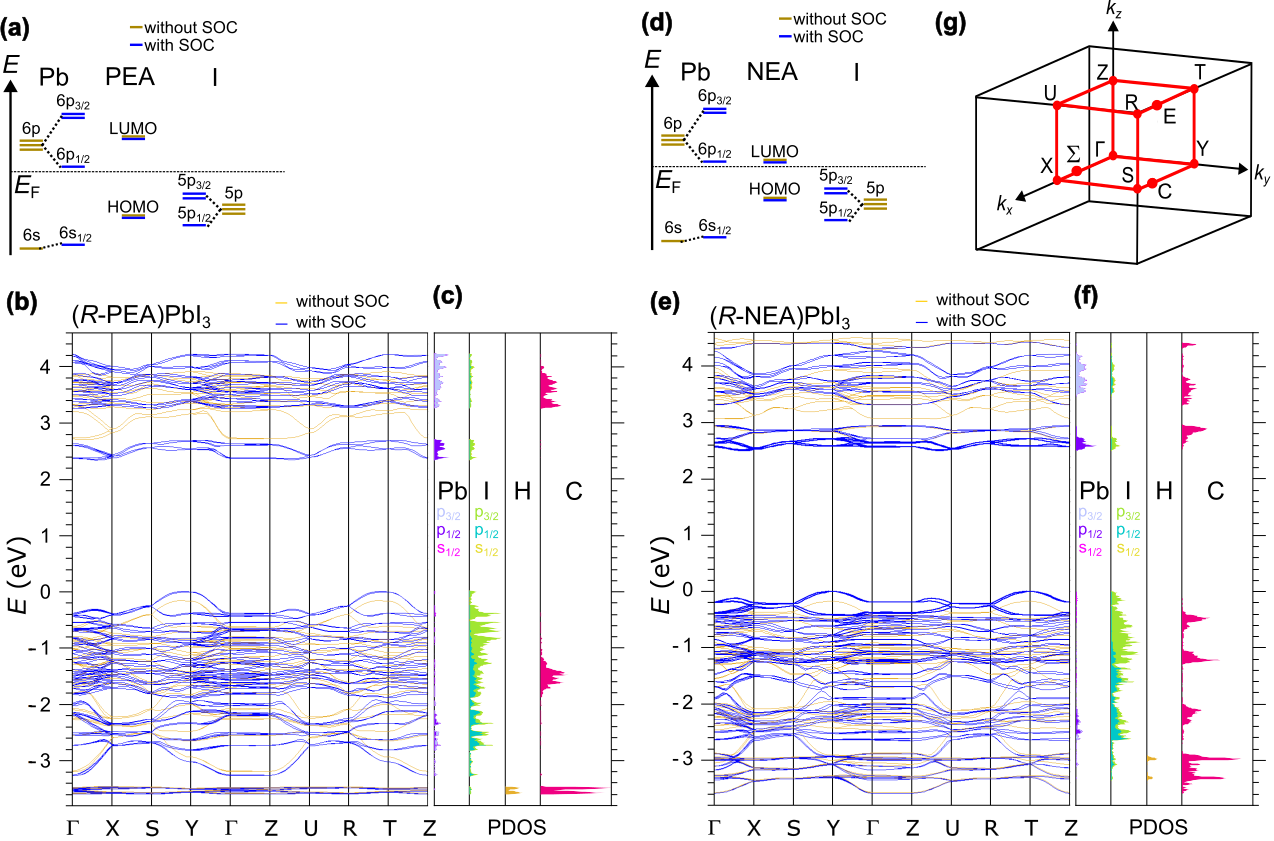}
  \caption{\label{fig:orbital}
    Electronic structures of (\textit{R}/\textit{S}-PEA)PbI$_3$ and (\textit{R}/\textit{S}-NEA)PbI$_3$.
    (a),(d) Schematic orbital characters near the band edges for the PEA and NEA compounds, respectively.
    (b),(e) Overall band structures of (\textit{R}-PEA)PbI$_3$ and (\textit{R}-NEA)PbI$_3$, respectively, calculated without SOC (orange) and with SOC (blue).
    (c),(f) Corresponding partial densities of states(PDOS).
    (g) Brillouin zone and the $k$-path used for the band-structure calculations.}
\end{figure*}
Density functional theory (DFT) calculations were performed using the \textit{Quantum ESPRESSO} package~\cite{QE2009,QE2017}.
The crystal structures used in the calculations were taken from experimental data obtained at $T = 100$~K.
The lattice constants $a$, $b$, and $c$ for (\textit{R}/\textit{S}-PEA)PbI$_3$ and (\textit{R}/\textit{S}-NEA)PbI$_3$ are summarized in Table~\ref{tab:lattice_constants}~\cite{SM}.
Calculations were carried out mainly for crystals constructed from the \textit{R}-type molecules.
For the exchange--correlation functional, we employed the generalized-gradient approximation (GGA) proposed by Perdew, Burke, and Ernzerhof (PBE)~\cite{Perdew2008}.
A plane-wave basis set within the projector augmented-wave (PAW) methodology was adopted. Both scalar-relativistic and fully relativistic pseudopotentials were used; in the main text, the corresponding calculations are referred to as ``without SOC'' and ``with SOC,'' respectively.
The valence-electron configurations were taken as $1s^1$ for H, $2s^22p^2$ for C, $2s^22p^3$ for N, $5s^25p^5$ for I, and $6s^26p^25d^{10}$ for Pb.
The plane-wave energy cutoffs were set to 50~Ry for the wavefunctions and 200~Ry for the charge density in both the (\textit{R}/\textit{S}-PEA)PbI$_3$ and the
(\textit{R}/\textit{S}-NEA)PbI$_3$ calculations.
A uniform $8\times8\times8$ $k$-point mesh was employed for the self-consistent-field loops.
For each compound, energies were referenced to the energy of the topmost valence band at the Y point obtained from the calculation with SOC.
For clarity, the sizes along the different directions in the Brillouin zone are shown on the same scale.

\section{\label{sec:overall}Overall Electronic Band Structure}
To understand the electronic states near the Fermi energy, we first examine the atomic and molecular orbitals that mainly contribute to the conduction and valence bands and discuss the role of SOC.
Figures~\ref{fig:orbital}(a) and \ref{fig:orbital}(d) schematically illustrate the relevant atomic and molecular orbitals near the Fermi energy for (\textit{R}/\textit{S}-PEA)PbI$_3$ and (\textit{R}/\textit{S}-NEA)PbI$_3$, respectively.
We first consider the case without SOC.
In both compounds, the valence bands are mainly derived from the I-$5p$ orbitals, and the highest occupied molecular orbital (HOMO) level of the chiral molecules is located well below the Fermi energy $E_{\rm F}$ and therefore does not directly contribute to the valence band near $E_{\rm F}$.
The conduction bands, however, differ between the two compounds.
In the absence of SOC, the conduction bands in (\textit{R}/\textit{S}-PEA)PbI$_3$ are primarily formed by the Pb-$6p$ orbitals, whereas those in (\textit{R}/\textit{S}-NEA)PbI$_3$ are mainly derived from the LUMO of the NEA molecule.
When SOC is included, the Pb-$6p$ and I-$5p$ orbitals split according to the total angular momentum $j$.
As a result, the character of the conduction bands is modified.
In (\textit{R}/\textit{S}-PEA)PbI$_3$, the conduction bands remain predominantly composed of the Pb-$6p$ orbitals, irrespective of the presence of SOC.
In contrast, in (\textit{R}/\textit{S}-NEA)PbI$_3$, the lowest conduction bands acquire a mixed character consisting of the Pb-$6p_{j=1/2}$ orbitals and the LUMO of the NEA molecule.
These properties are also reflected in the partial density of states (PDOS) resolved into the $j$ components of the atomic orbitals with SOC [Figs.~\ref{fig:orbital}(c) and \ref{fig:orbital}(f)].
Because the NEA molecule is larger than the PEA molecule, its HOMO--LUMO gap is reduced, which in turn leads to hybridization between the molecular LUMO and the lowest conduction bands mainly composed of Pb-$6p_{j=1/2}$ orbitals in (\textit{R}/\textit{S}-NEA)PbI$_3$.
This trend suggests that reducing the HOMO--LUMO gap---for example, by enlarging or substituting chiral molecules---provides an additional route to tuning the electronic states of chiral HOIPs.

We now turn to the overall band dispersions.
Figures~\ref{fig:orbital}(b) and \ref{fig:orbital}(e) show the calculated band structures of (\textit{R}-PEA)PbI$_3$ and (\textit{R}-NEA)PbI$_3$, respectively,
along the high-symmetry lines in the Brillouin zone [Fig.~\ref{fig:orbital}(g)].
Hereafter, the $a$, $b$, and $c$ axes correspond to the $x$, $y$, and $z$ axes, respectively.
In the presence of SOC (blue curves), both compounds have their conduction-band minimum along the $\Gamma$--X line, $(k_x, 0, 0)$, and their valence-band maximum near the Y point, $(0, \pi/b, 0)$.
It should be noted that the conduction-band minimum is located far from the high-symmetry points, indicating that a minimal nearest-neighbor model is insufficient.
The calculated indirect band gaps are 2.3~eV for (\textit{R}-PEA)PbI$_3$ and 2.5~eV for (\textit{R}-NEA)PbI$_3$.
Although band gaps are generally underestimated in first-principles calculations, the larger gap of the NEA-based compound is consistent with experimentally estimated optical gaps, approximately 2.63 eV for (\textit{R}/\textit{S}-PEA)PbI$_3$~\cite{Makhija2025} and 3 eV for (\textit{R}/\textit{S}-NEA)PbI$_3$~\cite{Ishii2025}.
Regarding the dimensionality of the electronic structure, both compounds exhibit negligible dispersion along the $k_z$ direction.
The band structures along the $k_z = 0$ ($\Gamma$-X-S-Y-$\Gamma$) and $k_z = \pi/c$ (Z-U-R-T-Z) lines are nearly identical, indicating that these systems are quasi-two-dimensional despite the presence of one-dimensional chains in their crystal structure.
To further examine the influence of SOC on the energy dispersions, we also show, for comparison, the band structures calculated without SOC as orange lines in Figs.~\ref{fig:orbital}(b) and \ref{fig:orbital}(e).
The most pronounced effect of SOC in both the PEA- and NEA-based compounds is the splitting of the Pb-derived conduction bands, as mentioned above.
In contrast, SOC has only a minor impact on the energy levels of the molecular LUMO, owing to the weak atomic SOC of the elements C, N, and H.
Consequently, in (\textit{R}/\textit{S}-NEA)PbI$_3$, the lowest conduction band is composed solely of the molecular LUMO in the absence of SOC~\cite{Xiao2024}, whereas including SOC leads to the formation of the lowest conduction bands in which the Pb orbitals become dominant and hybridize with the LUMO.

\section{\label{sec:spinsplit}Spin-Split Bands}
\begin{figure*}[!t]
  \includegraphics[width=17cm,clip]{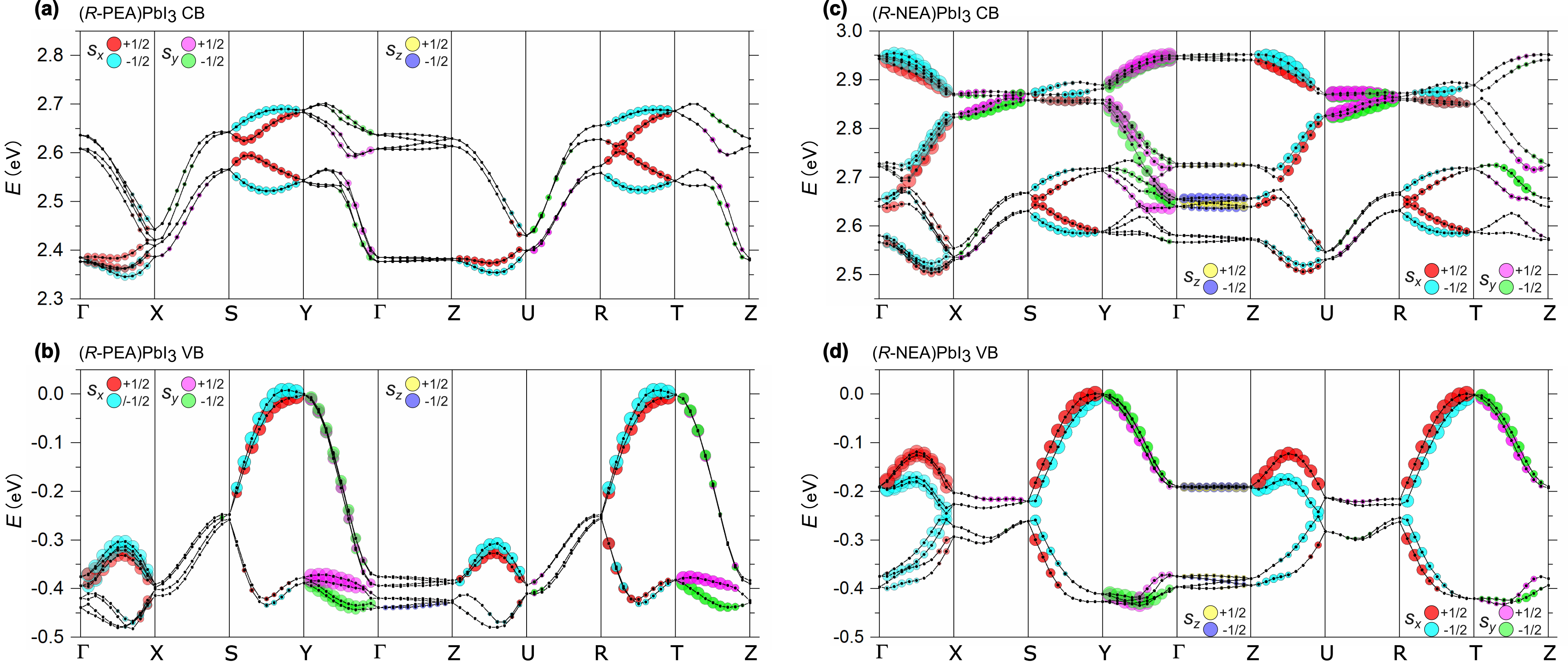}
  \caption{\label{fig:spinsplit}
    Spin-split band dispersions of (\textit{R}-PEA)PbI$_3$ and (\textit{R}-NEA)PbI$_3$.
    (a),(b) Conduction-band (CB) and valence-band (VB) dispersions of (\textit{R}-PEA)PbI$_3$, respectively; (c),(d) corresponding CB and VB dispersions of (\textit{R}-NEA)PbI$_3$.
    Colored circles represent the spin-polarization components $s_x$, $s_y$, and $s_z$.
    The sign convention for each component is indicated by the legends, where $\pm1/2$ corresponds to full spin polarization.
    The circle radius is proportional to the magnitude of the corresponding spin polarization.}
\end{figure*}
Next, we examine the spin splitting of the energy bands.
Figure~\ref{fig:spinsplit} shows the conduction and valence bands of (\textit{R}-PEA)PbI$_3$ and (\textit{R}-NEA)PbI$_3$ near the Fermi energy.
Along a symmetry axis in $k$-space, the spin expectation value at a given $k$ point, obtained by averaging over all degenerate states when degeneracy is present, has a finite component only along the axis.
Indeed, along the $\Gamma$--X line, only the $s_x$ component is finite, whereas along the Y--$\Gamma$ line, only the $s_y$ component is finite.
This parallel spin texture, in which the spin polarization is parallel to the corresponding $k$-space direction, is a characteristic feature of
nonpolar chiral systems and stands in contrast to the circular spin texture of Rashba (polar) systems~\cite{Wang2020}. 
Note that the magnitude of the spin polarization depends on the wave vector.
There are directions in which a wide region of almost complete spin polarization appears, such as in the topmost valence band along the S--Y line; in contrast, other directions, such as the X--S line, exhibit only small spin polarization.
Furthermore, there are directions in which the spin polarization is large but the energy splitting between opposite-spin states is small, as observed for the Y--$\Gamma$ line of the topmost valence band in (\textit{R}-PEA)PbI$_3$.

\begin{figure}[!t]
  \includegraphics[width=5cm,clip]{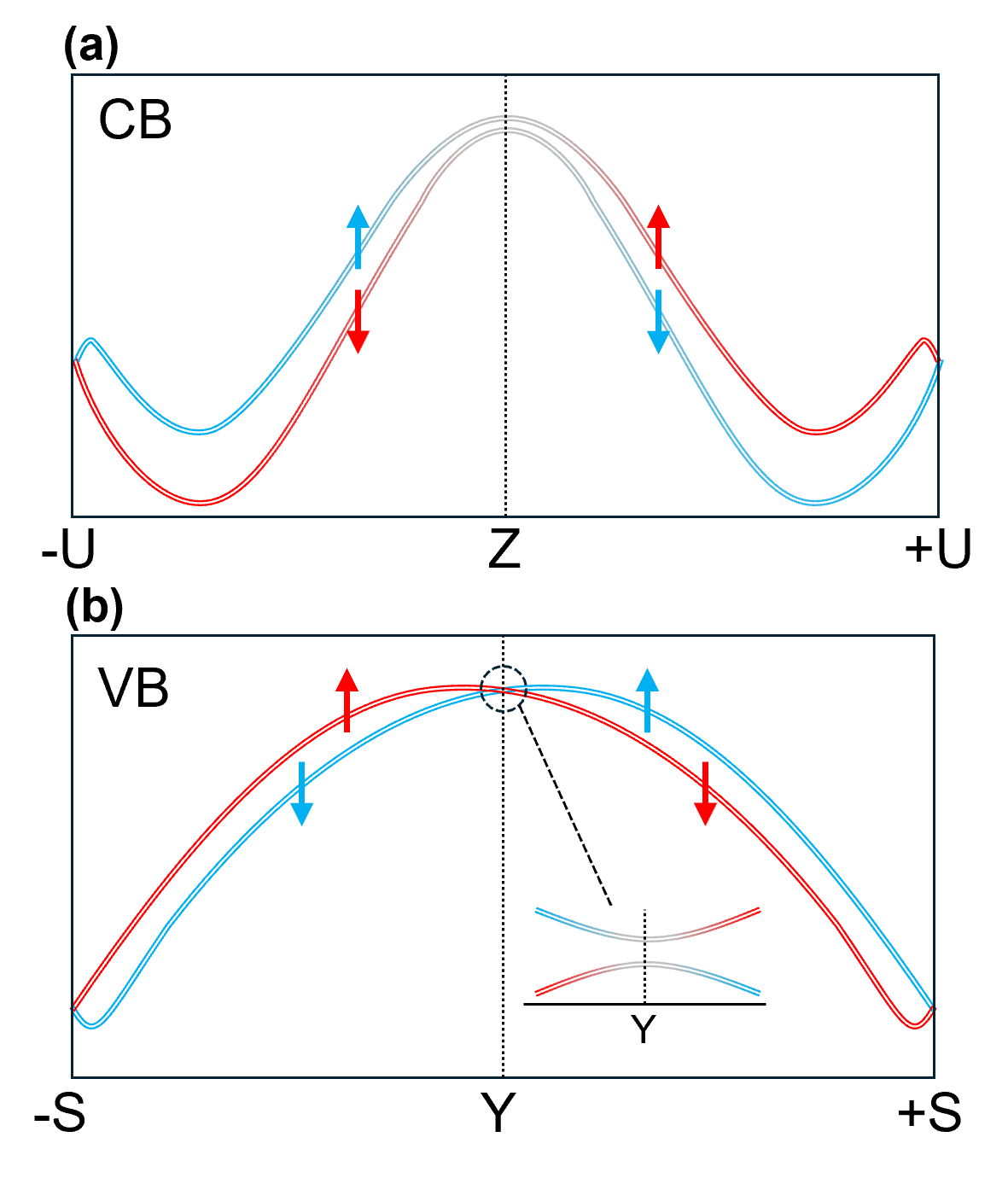}
  \caption{\label{fig:schematic}
    Schematic spin-split band dispersions of the lowest conduction band (a) and the highest valence band (b).
    Red and blue branches represent opposite spin polarizations, while gray branches denote zero spin polarization.
    The spin polarization gradually vanishes toward the Y and Z points.}
\end{figure}

Interestingly, both the valence-band maximum along the S--Y (R--T) line and the conduction-band minimum along the $\Gamma$--X (Z--U) line exhibit large spin splitting.
These large spin splittings may be related to the large CD signals observed experimentally~\cite{Chen2019,Ishii2020}.
Comparing the $\Gamma$--X and Z--U lines, one finds that the number of spin-split bands differs by a factor of two.
This is because nonsymmorphic crystal symmetry enforces double degeneracy at the Brillouin-zone boundary, an effect known as the band sticking effect~\cite{KonigMermin2000}.
To understand the essential features of the spin-split bands of the two materials, we show schematic spin-split bands along the S--Y line for the highest valence bands and along the U--Z line for the lowest conduction bands in Fig.~\ref{fig:schematic}.
Note that all branches shown are doubly degenerate, and the two degenerate states have the same spin component parallel to the corresponding $k$-space line.
Note also that the conduction bands along the $\Gamma$--X line can be regarded as resulting from an additional splitting of the bands along the U--Z line; thus, we focus on the bands along the U--Z line for simplicity.
At first glance, the valence bands appear to be typical spin-split bands with a linear dispersion, as in Rashba systems.
However, as discussed in the Section~\ref{sec:grouptheory}, a tiny but finite band gap opens at the Y point, and the spin polarization approaches zero toward this point.
The conduction bands show this unconventional feature more clearly around the Z point.

In the band structure of (\textit{R}-NEA)PbI$_3$, the bands located above approximately 2.65~eV, which are derived primarily from the LUMO of the NEA molecule, also exhibit large spin polarization.
This indicates that hybridization between Pb/I orbitals and LUMO-derived states induces large spin polarization even in the molecular-orbital-derived bands.
More broadly, these findings reveal that the organic molecules, previously regarded as merely providing a structural framework, can play an active role in determining the electronic properties.

\begin{figure*}[!t]
  \includegraphics[width=17cm,clip]{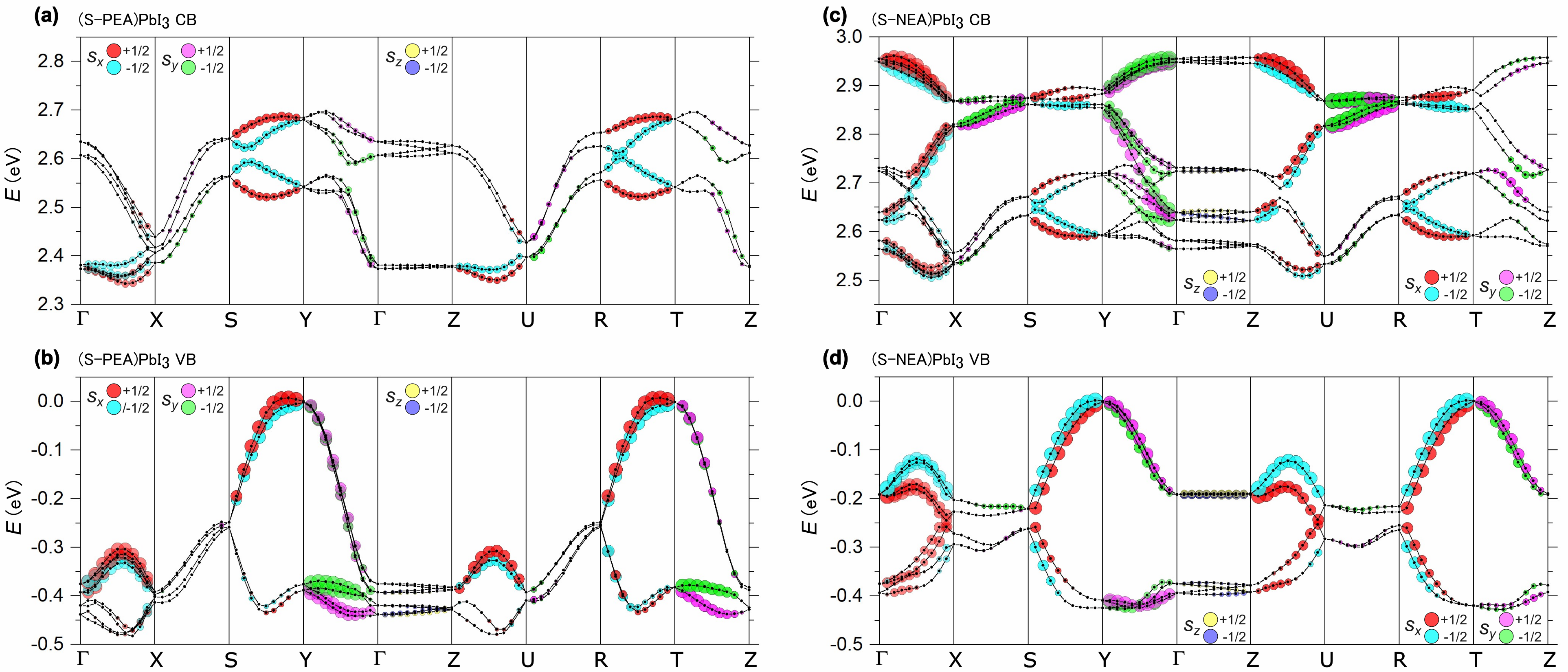}
  \caption{\label{fig:stype}
    Spin-split band dispersions of (\textit{S}-PEA)PbI$_3$ and (\textit{S}-NEA)PbI$_3$.
    (a),(b) Conduction-band (CB) and valence-band (VB) dispersions of (\textit{S}-PEA)PbI$_3$, respectively; (c),(d) corresponding CB and VB
    dispersions of (\textit{S}-NEA)PbI$_3$.
    Colored circles represent the spin-polarization components $s_x$, $s_y$, and $s_z$.
    The sign convention for each component is indicated by the legends, where $\pm1/2$ corresponds to full spin polarization.
    The circle radius is proportional to the magnitude of the corresponding spin polarization.}
\end{figure*}

Figure ~\ref{fig:stype} shows the spin-split band structures of the \textit{S}-type compounds. While their energy dispersions are nearly identical to those of the corresponding \textit{R}-type compounds, our calculations confirm that the spin polarizations are reversed between the \textit{R}- and \textit{S}-type compounds.
This behavior is consistent with their enantiomorphic relationship.

\section{\label{sec:signsCD}Relationship Among the Signs of Spin Texture, Structural Chirality, and CD Signals}

A comparison of the spin polarization between the PEA- and NEA-based compounds with the same molecular handedness, either \textit{R} or \textit{S}, reveals that the signs of the spin texture reversed over almost the entire Brillouin zone.
In both systems, the [PbI$_6$]$^{4-}$ octahedra acquire chiral distortions induced by the chiral molecules, giving rise to electronic chirality.
Therefore, the opposite spin textures in $k$-space suggest that the [PbI$_6$]$^{4-}$ octahedra in (\textit{R}-PEA)PbI$_3$ and (\textit{R}-NEA)PbI$_3$ [(\textit{S}-PEA)PbI$_3$ and (\textit{S}-NEA)PbI$_3$] have opposite chiral distortions.
To support this interpretation from a structural viewpoint, we evaluate the structural chiral distortions of the [PbI$_6$]$^{4-}$ octahedra using an empirical method based solely on the atomic coordinates for the two \textit{R}-type materials.
For each face of an [PbI$_6$]$^{4-}$ octahedron, defined by three Pb$\to$I vectors $\bm{r}_i$, $\bm{r}_j$, and $\bm{r}_k$, we construct the scalar triple product $\chi_{ijk} = \bm{r}_i \cdot (\bm{r}_j \times \bm{r}_k)$ [Fig.~\ref{fig:structure}(e)].
This quantity is time-reversal-even and spatial-inversion-odd, and thus possesses the same symmetry as chirality~\cite{Barron2020}.
By calculating this scalar triple product for all eight faces with a consistent counterclockwise convention relative to the face normal and summing the results [Fig.~\ref{fig:structure}(e)], we define an indicator $\chi$ of the structural chiral distortion of the octahedron.
This indicator takes the values $-265.3$~\AA$^3$ for (\textit{R}-PEA)PbI$_3$ and $+243.7$~\AA$^3$ for (\textit{R}-NEA)PbI$_3$, demonstrating a sign reversal between the two \textit{R}-type materials.
These opposite signs of both electronic and structural chirality in the PEA- and NEA-based materials with the same molecular handedness are consistent with the experimentally observed opposite CD signals between the two materials~\cite{Chen2019,Ishii2020}.

\section{\label{sec:origin}Origin of Chirality Transfer}

Here, we demonstrate that (\textit{R}-NEA)PbI$_3$ serves as a counterexample to the explanation proposed by Wei \textit{et al.}, in which the origin of the spin texture is attributed to the relative displacement between the centers of positive and negative charges in the [PbI$_6$]$^{4-}$ octahedra along the $z$ direction.
In the case of (\textit{R}-PEA)PbI$_3$, the displacements are $\Delta x = -0.014$~\AA, $\Delta y = -0.023$~\AA, and $\Delta z = 0.152$~\AA, indicating that the displacement vector is indeed primarily along the $z$ direction.
In contrast, in (\textit{R}-NEA)PbI$_3$, the displacements are $\Delta x = 0.059$~\AA, $\Delta y = 0.004$~\AA, and $\Delta z = 0.012$~\AA, showing no pronounced displacement along the $z$ direction.
Nevertheless, the spin splitting is comparable or even larger in (\textit{R}-NEA)PbI$_3$ than in (\textit{R}-PEA)PbI$_3$, as discussed in the Section~\ref{sec:comparison}.
This result indicates that the charge displacement of the octahedron---namely, the electric polarization of the octahedron in the classical sense---cannot directly account for the magnitude of the spin splitting.
This is reasonable given that the present system is chiral, and polarity alone cannot capture the magnitude of chirality.
In other words, the spin splitting in chiral systems cannot be explained solely in terms of polar distortions.

\section{Comparison of Spin Splitting Between (\textit{R}-PEA)P\lowercase{b}I$_3$ and (\textit{R}-NEA)P\lowercase{b}I$_3$}
\label{sec:comparison}

\begin{figure}[!t]
  \includegraphics[width=8.6cm,clip]{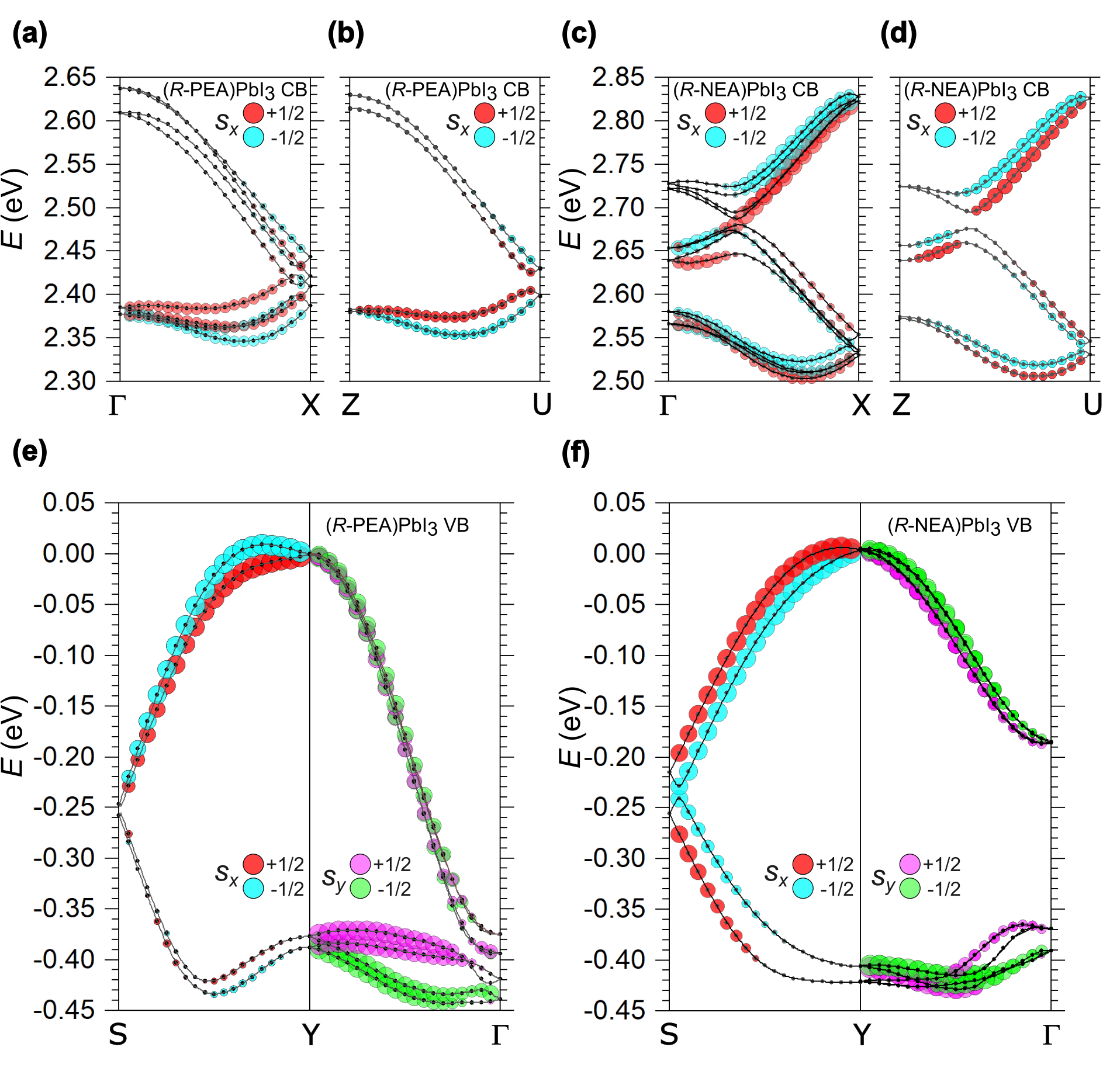}
  \caption{\label{fig:enlarged}
    Enlarged views of the spin-split bands of (\textit{R}-PEA)PbI$_3$ and (\textit{R}-NEA)PbI$_3$.
    (a),(b) Conduction bands of (\textit{R}-PEA)PbI$_3$ near the $\Gamma$--X and Z--U lines, respectively, and (c),(d) corresponding conduction bands of (\textit{R}-NEA)PbI$_3$.
    (e),(f) Valence bands of (\textit{R}-PEA)PbI$_3$ and (\textit{R}-NEA)PbI$_3$, respectively.
    Colored circles represent the spin-polarization components $s_x$ and $s_y$.
    The legends indicate the fully polarized values, $\pm1/2$, for each component, and the circle radius scales linearly with the magnitude of the corresponding spin polarization.}
\end{figure}

Next, we discuss details of the spin-splitting properties.
Figure~\ref{fig:enlarged} shows enlarged band structures near the band edges.
Figures ~\ref{fig:enlarged}(a)--\ref{fig:enlarged}(d) show the conduction bands along the $\Gamma$--X and Z--U lines, while Fig.~\ref{fig:enlarged}(e) and Fig.\ref{fig:enlarged}(f) show the valence bands along the S--Y--$\Gamma$ line.

\begin{figure}[!t]
  \includegraphics[width=8.6cm,clip]{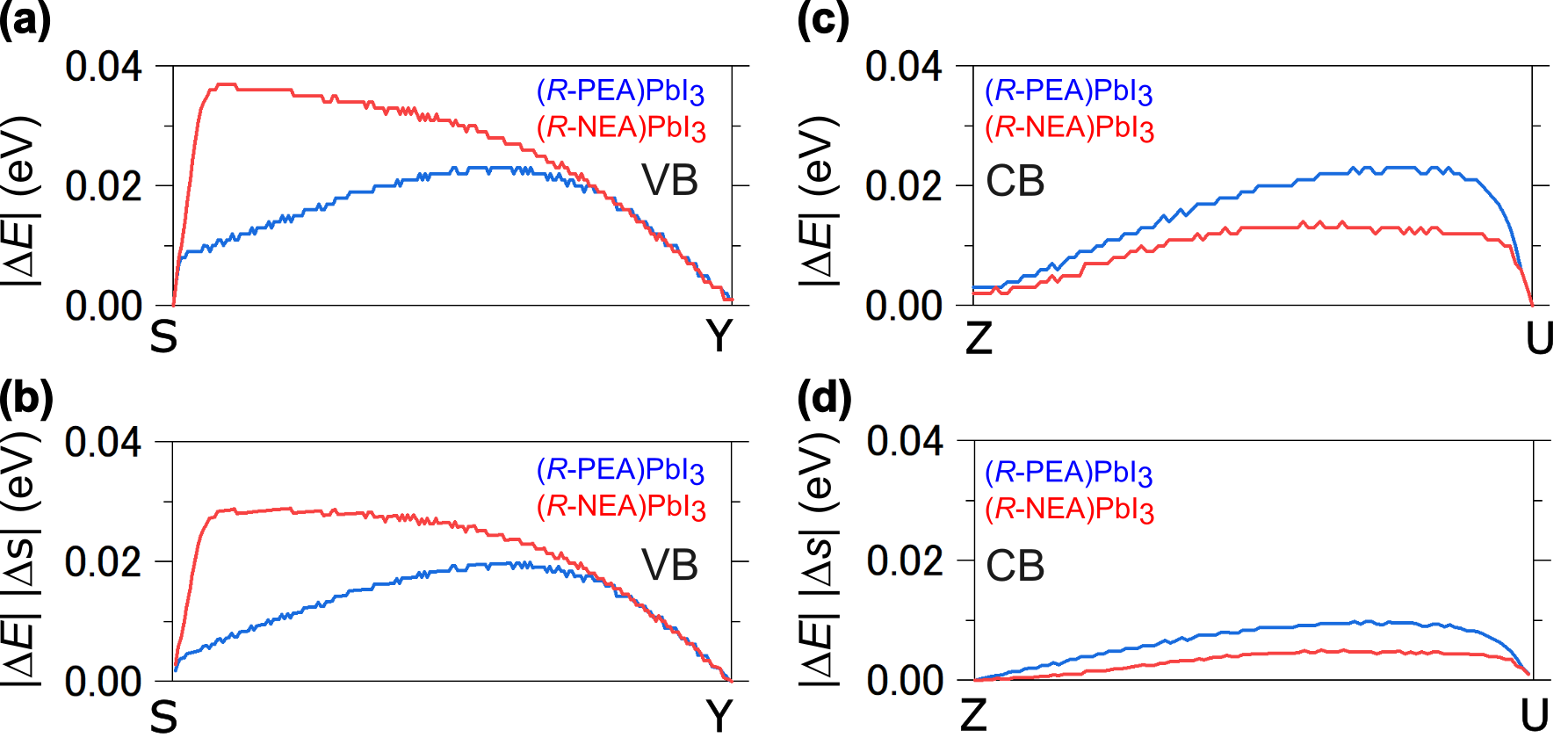}
  \caption{\label{fig:splitting}
    Energy splitting magnitude $|\Delta E|$ and the effective spin splitting magnitude $|\Delta E||\Delta s|$ for the valence and conduction bands of (\textit{R}-PEA)PbI$_3$ and (\textit{R}-NEA)PbI$_3$.
    (a) $|\Delta E|$ and (b) $|\Delta E||\Delta s|$ along the S--Y line.
    (c) $|\Delta E|$ and (d) $|\Delta E||\Delta s|$ along the Z--U line.
    Here, $\Delta E$ denotes the energy difference between two adjacent doubly degenerate band pairs, and $\Delta s$ the difference in spin polarization between the two pairs.}
\end{figure}

We first focus on the valence bands.
Near the valence-band maximum along the S--Y--$\Gamma$ line [Figs.~\ref{fig:enlarged}(e) and \ref{fig:enlarged}(f)], a clear spin splitting is observed.
Along the S--Y line on the zone boundary, each band is doubly degenerate owing to the band-sticking effect; therefore, the eight plotted valence bands appear as four distinct branches.
In the following discussion, we treat the doubly degenerate states on the zone boundary as pairs and define the energy splitting $\Delta E$ as the energy difference between two adjacent band pairs closest to the band edge.
Near the S and Y points, the energy splitting $\Delta E$ increases linearly with the wave-vector deviation $\Delta k_x$, measured from the high-symmetry points, following $\Delta E \sim 2\alpha \Delta k_x$, where $\alpha$ is the linear coefficient of energy splitting.
To discuss the splitting quantitatively, we plot the absolute energy splitting $|\Delta E|$ between adjacent band pairs along the S--Y line in Fig.~\ref{fig:splitting}(a).
Near the Y point, the linear coefficients $\alpha_{\rm Y}$ are estimated to be approximately 0.1~eV\AA\ for both (\textit{R}-PEA)PbI$_3$ and (\textit{R}-NEA)PbI$_3$.
Notably, the corresponding coefficients near the S point, $\alpha_{\rm S}$, are estimated to be approximately 0.9~eV\AA\ for both compounds, close to 1~eV\AA\ and comparable to those of strong Rashba systems~\cite{Acosta2020}.
Note that the large linear coefficients near the S point are not induced by SOC, but instead originate from a linear band structure that already exists in the absence of SOC, as discussed in the Section~\ref{sec:anticrossing}.
This result suggests that the two materials have spin-dependent intrachain hopping parameters of similar magnitude along the [PbI$_6$]$^{4-}$ octahedra chain direction.

We next focus on the intermediate region of the S--Y line away from the high-symmetry points.
Notably, although the coefficients $\alpha_{\rm Y}$ and $\alpha_{\rm S}$ near the Y and S points are similar for the two compounds, the splitting in (\textit{R}-NEA)PbI$_3$ remains larger than that in (\textit{R}-PEA)PbI$_3$ over a wide region of the S--Y line.
This enhanced spin splitting in the intermediate region can be attributed to the band (anti)crossing discussed later.
To characterize the spin splitting properly, it is necessary to consider not only the energy separation but also the spin polarization of the bands.
As noted above, the spin polarization of an energy band can vary with the wave vector owing to the multiband nature of the present chiral HOIPs.
Thus, a large $\Delta E$ does not necessarily indicate a large effective spin splitting, which requires both a large $\Delta E$ and a large difference in spin polarization between the corresponding bands.
Therefore, we plot the quantity $|\Delta E|\cdot|\Delta s|$ in Fig.~\ref{fig:splitting}(b), where $\Delta s$ denotes the difference in the spin expectation values along the symmetry axis between the corresponding band pairs.
For the valence bands along the S--Y line, $|\Delta E|\cdot|\Delta s|$ shows behavior similar to that of $|\Delta E|$, indicating that the effective spin splitting of (\textit{R}/\textit{S}-NEA)PbI$_3$ is indeed larger than that of (\textit{R}/\textit{S}-PEA)PbI$_3$.
Along the Y--$\Gamma$ line, which is parallel to the interchain axis $b$, (\textit{R}-PEA)PbI$_3$ exhibits a large spin polarization at each $k$ point but only a small energy splitting between opposite-spin states [Fig.~\ref{fig:enlarged}(e)].
In (\textit{R}-NEA)PbI$_3$, however, the spin splitting is larger [Fig.~\ref{fig:enlarged}(f)].
This result suggests that interchain spin-dependent hopping is stronger in (\textit{R}/\textit{S}-NEA)PbI$_3$.

Next, we turn to the conduction bands [Figs.~\ref{fig:enlarged}(a)--\ref{fig:enlarged}(d)].
Large spin splitting also appears around the conduction-band minimum.
The band structure along the Z--U line on the zone boundary $(k_x, 0, \pi/c)$ is simpler than that along the $\Gamma$--X line $(k_x, 0, 0)$ owing to the band-sticking effect; thus, we evaluate the spin splitting of the conduction bands along the Z--U line for simplicity.
The energy difference $|\Delta E|$ between adjacent band pairs along the Z--U line for (\textit{R}-PEA)PbI$_3$ and (\textit{R}-NEA)PbI$_3$, shown in Fig.~\ref{fig:splitting}(c), indicates that the magnitude of the spin splitting near the high-symmetry points is similar in the two systems, as in the case of the topmost valence bands.
This result again implies that the magnitudes of the spin-dependent intrachain hopping are comparable in (\textit{R}/\textit{S}-NEA)PbI$_3$ and (\textit{R}/\textit{S}-PEA)PbI$_3$.
In the intermediate region of the Z--U line, $|\Delta E|$ appears to be larger in (\textit{R}-PEA)PbI$_3$.
When the quantity $|\Delta E|\cdot|\Delta s|$, which takes into account the degree of spin polarization, is considered, the effective spin splitting in the conduction bands is reduced by the incomplete spin polarization compared with the magnitude expected from $|\Delta E|$ alone.
This contrasts to the case of the edge of the valence bands, in which spins are nearly fully polarized.
Therefore, the effective spin splitting is larger at the valence-band edge than at the conduction-band edge, and this valence-band-edge splitting is more pronounced in (\textit{R}/\textit{S}-NEA)PbI$_3$ than in (\textit{R}/\textit{S}-PEA)PbI$_3$.

\section{\label{sec:anticrossing}Enhancement of Spin Splitting Due to
  Band (Anti)crossing}

In the following, we discuss how band (anti)crossing enhances spin splitting.
Here, the distinction between band crossing and anticrossing is not essential to the following discussion.
As noted above, in the valence bands, the linear coefficients of the spin splitting near the S and Y points are nearly identical for (\textit{R}-PEA)PbI$_3$ and (\textit{R}-NEA)PbI$_3$, whereas the magnitude of the spin splitting in the intermediate region along the S--Y line is significantly larger in (\textit{R}-NEA)PbI$_3$ than in (\textit{R}-PEA)PbI$_3$ [Fig.~\ref{fig:splitting}(a)].
In typical discussions of spin-split bands, the magnitude of the linear coefficient is often regarded as the primary factor determining the size of the spin splitting.
However, the present result shows that even when the linear coefficients near high-symmetry points are nearly the same, the spin splitting away from these points can differ substantially.
Understanding the origin of this difference is important for identifying mechanisms that enhance spin splitting over a wide region of the Brillouin zone.

\begin{figure}[!t]
  \includegraphics[width=8.6cm,clip]{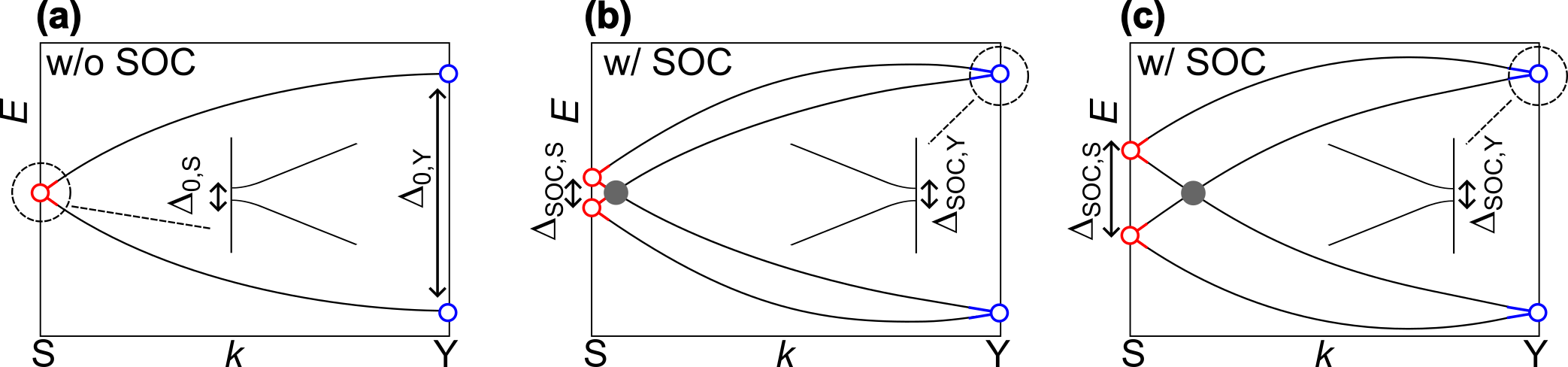}
  \caption{\label{fig:schematic_anti}
    Schematic band dispersions showing the enhancement of spin splitting induced by band (anti)crossing.
    Four-orbital bands are shown along the S--Y line.
    The black lines represent bands that are doubly degenerate due to the band-sticking effect.
    (a) Bands without SOC, where spin degrees of freedom are not included.
    (b),(c) Bands with SOC included for the cases of small and large SOC-induced gaps $\Delta_{\rm SOC,S}$, respectively.}
\end{figure}

To illustrate this point, we consider schematic four-orbital valence bands along the S--Y line, as shown in Fig.~\ref{fig:schematic_anti}.
Figure~\ref{fig:schematic_anti}(a) shows the bands without SOC, while Figs.~\ref{fig:schematic_anti}(b) and \ref{fig:schematic_anti}(c) show the corresponding bands with SOC included.
In the absence of SOC, when spin degrees of freedom are not considered, both the S and Y points host two doubly degenerate states, and the bands along the S--Y line are doubly degenerate owing to the band-sticking effect.
We denote the gaps between the two degenerate pairs at the Y and S points without SOC by $\Delta_{0,\rm Y}$ and $\Delta_{0,\rm S}$, respectively.
At the S point, based on the calculation results, we assume that $\Delta_{0,\rm S}$ is negligibly small, so that a linear dispersion with coefficient $\alpha_{0,\rm S}$ around the S point can be defined even in the spinless system, whereas no linear dispersion is present near the Y point owing to the large $\Delta_{0,\rm Y}$.
Upon including SOC, the fourfold degeneracies at the S and Y points, including spin degrees of freedom, are split into two doubly degenerate pairs separated by the SOC-induced gaps $\Delta_{\rm SOC,S}$ and $\Delta_{\rm SOC,Y}$, respectively.
SOC also introduces linear-in-$\bm{k}$ contributions characterized by $\alpha_{\rm SOC,S}$ and $\alpha_{\rm SOC,Y}$, which are proportional to the SOC strength, near these points.
The resulting linear coefficients are therefore given by $\alpha_{\rm S} \sim \alpha_{0,\rm S} + \alpha_{\rm SOC,S}$ and $\alpha_{\rm Y} \sim \alpha_{\rm SOC,Y}$.
Based on the calculation results presented above, we take $\Delta_{\rm SOC,S} \gg \Delta_{0,\rm S}$ and $|\alpha_{\rm S}| \sim |\alpha_{0,\rm S}| \gg |\alpha_{\rm SOC,S}|$, and $\Delta_{\rm SOC,Y}$ is taken to be negligibly small.
The cases shown in Figs.~\ref{fig:schematic_anti}(b) and \ref{fig:schematic_anti}(c) can be associated with (\textit{R}-PEA)PbI$_3$ and (\textit{R}-NEA)PbI$_3$, respectively.
The two cases have nearly identical linear coefficients, $|\alpha_{\rm S}| \sim |\alpha_{0,\rm S}|$ and $|\alpha_{\rm Y}| \sim |\alpha_{\rm SOC,Y}|$, but differ significantly in the SOC-induced gap $\Delta_{\rm SOC,S}$.
Along the S--Y line, the splitting in Fig.~\ref{fig:schematic_anti}(c) is larger than that in Fig.~\ref{fig:schematic_anti}(b) owing to the large
$\Delta_{\rm SOC,S}$ and the associated band (anti)crossing.

In the valence bands considered here, owing to the nonsymmorphic crystal symmetry, the two pairs of the spin-split branches emerging from the S point along the S--Y path switch partners and become degenerate again at the Y point, giving rise to a band (anti)crossing.
As a result, the energy separation between adjacent spin-split band pairs is enhanced near the (anti)crossing point.
These considerations indicate that the spin splitting in the intermediate region of a symmetry line is determined not only by the linear coefficients near the high-symmetry points but also by the combined effects of band (anti)crossing and the SOC-induced gap at the high-symmetry points.
If the experimentally observed difference in the CD signals of (\textit{R}/\textit{S}-PEA)PbI$_3$ and (\textit{R}/\textit{S}-NEA)PbI$_3$ originates from the difference in spin splitting in the intermediate region of the valence bands, its origin should therefore be attributed not to the linear coefficients but to the SOC-induced gap at the high-symmetry point.
Because the enhancement of spin splitting associated with band (anti)crossing is inherently a multiband effect, chiral HOIPs, which naturally host multiple relevant bands near the band edges, provide a natural platform for investigating how band hybridization and SOC-induced gaps at high-symmetry points cooperatively determine spin splitting.
Note that a similar relationship between spin splitting and band (anti)crossing has also been discussed in polar systems, where large Rashba coefficients can arise from such (anti)crossing behavior~\cite{Acosta2020}.

\section{\label{sec:grouptheory}Group-Theoretical Considerations}

Here, based on group-theoretical arguments, we show that nonsymmorphic symmetries give rise to nontrivial band degeneracies and spin splittings in the present chiral HOIPs; these effects have not been explicitly addressed in previous studies.
First, we briefly review general properties common to systems belonging to nonsymmorphic space groups.
Energy eigenstates at a given wave vector $\bm{k}$ belong to one of the irreducible representations (irreps) of the little group at $\bm{k}$.
If the corresponding irrep is two-dimensional or higher, degeneracy necessarily occurs at the $k$-point.
When the space group is symmorphic, or when the space group is nonsymmorphic but $\bm{k}$ does not lie on the Brillouin-zone boundary, the irreps of the little group at $\bm{k}$ can be expressed in terms of the irreps of the crystal point group associated with the space group (e.g.\ $222$ from $P2_12_12_1$).
In contrast, when the space group is nonsymmorphic and $\bm{k}$ lies on the Brillouin-zone boundary, irreps different from those of the corresponding crystal point group can appear.
In other words, crystal structures belonging to nonsymmorphic space groups can exhibit nontrivial degeneracy patterns on the Brillouin-zone boundary, for instance, the band-sticking effect~\cite{KonigMermin2000}.
In symmorphic systems, for high-symmetry points invariant under the point-group operations of the crystal, the degeneracy pattern at a high-symmetry point on the zone boundary is identical to that at the origin of $k$-space (the $\Gamma$ point).
In contrast, in nonsymmorphic systems, additional degeneracies or lifting of degeneracies can occur at high-symmetry points on the zone boundary.

\begin{figure}[!t]
  \includegraphics[width=8.6cm,clip]{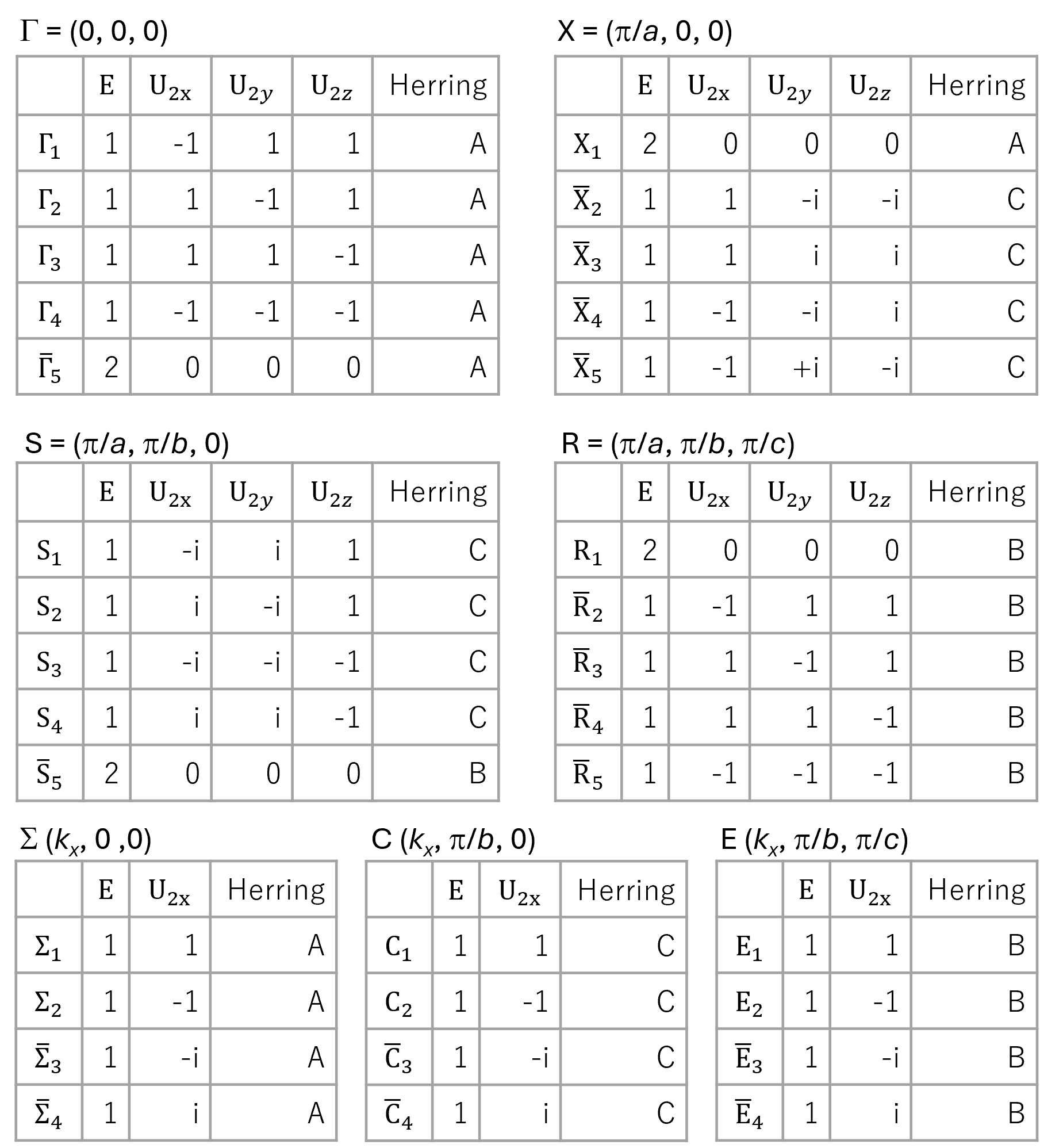}
  \caption{\label{fig:chartable}
    Character tables of the space group $P2_12_12_1$ at the high-symmetry points and along the high-symmetry lines.
    Irreducible representations without overbars denote single-valued representations (without SOC), whereas those with overbars denote double-valued representations (with SOC).
    $E$ is the identity operation, and $U_{2x}$, $U_{2y}$, and $U_{2z}$ denote the screw operations along the corresponding helical axes.
    The Herring criterion labels A, B, and C indicate real, pseudoreal, and complex representations, respectively.}
\end{figure}

We next apply this argument to the chiral HOIPs considered here, which belong to the nonsymmorphic space group $P2_12_12_1$.
The character tables of the spinful double-valued irreducible representations at each high-symmetry point for systems with $P2_12_12_1$ symmetry are shown in Fig.~\ref{fig:chartable}, together with the indices defined by Herring's criterion~\cite{Herring1937}, which diagnose whether time-reversal symmetry causes an additional degeneracy.
In $P2_12_12_1$, while the three crystallographic axes are not symmetry-equivalent, each axis possesses a twofold screw symmetry.
Consequently, the band-degeneracy structures follow the same pattern along the three directions.
For example, the high-symmetry points X, Y, and Z have the same degeneracy structure.
Therefore, it is sufficient to analyze the representative $k$ points $\Gamma$, X, S, and R as well as the points $\Sigma$ on $\Gamma$--X, C on Y--S, and E on T--R\@.

First, we consider symmetry-enforced degeneracies at the high-symmetry points.
These points are invariant under three mutually orthogonal twofold screw operations along the $x$, $y$, and $z$ directions.
At each high-symmetry point, we consider an energy eigenstate whose spin polarization is aligned with one of the three screw axes, for simplicity.
In this case, the spin polarization can be reversed by a twofold screw operation whose screw axis is perpendicular to the spin direction.

At the $\Gamma$ point, only the two-dimensional real representation $\Gamma_5$ appears.
One of the doubly degenerate states with finite spin polarization is transformed into the other state with opposite spin polarization under a
twofold screw operation whose screw axis is perpendicular to the spin direction.
Thus, the two degenerate states generally carry finite spin polarization, allowing spin-polarized bands to appear near the $\Gamma$ point.
According to Herring's criterion, time-reversal symmetry also protects this degeneracy but does not generate an additional degeneracy.

In contrast, at the X (Y or Z) point, one-dimensional complex representations appear.
This means that any of the three twofold screw operations maps an energy eigenstate not to its Kramers partner but to itself.
In addition, as mentioned above, a twofold screw operation whose axis is perpendicular to the spin direction flips the spin polarization.
These two facts force the spin polarization to vanish at the X point.
As a result, unconventional spin-split bands whose spin polarization vanishes as the X point is approached can appear.
At the X point, time-reversal symmetry alone enforces a twofold degeneracy forming a Kramers pair.

At the S point, a fourfold degeneracy emerges, owing to the three twofold screw symmetries together with time-reversal symmetry, and conventional spin-polarized bands can appear near the S point.
At the R point, a twofold degeneracy emerges, protected solely by time-reversal symmetry, and spin polarization approaches zero toward the R point, as at the X point.
In the band calculations for the present chiral HOIPs, the number of degenerate states indeed differs from one high-symmetry point to another, as described above.
In addition, the spin polarization is observed to approach zero toward the high-symmetry points on the zone boundary, except at the S, T, and U points.
However, near the valence-band maximum around the Y point, the splitting between the two doubly degenerate bands appears unresolved because the splitting is quantitatively very small; nevertheless, a tiny but finite splitting is actually realized, as schematically shown in Fig.~\ref{fig:schematic}.

Next, we turn to high-symmetry lines connecting high-symmetry points.
At the $\Sigma$ point $(k_x, 0, 0)$ along the $\Gamma$--X line, the (unitary) symmetry operation that leaves $\bm{k}$ invariant is only the screw operation $2_x$ combined with a translation.
Consequently, only one-dimensional representations are allowed.
The indices for Herring's criterion show that these representations are real, so time-reversal symmetry does not induce additional degeneracy.
Therefore, all band degeneracies are generically lifted on the $\Gamma$--X line.
In contrast, states at the C point $(k_x, \pi/b, 0)$ on the Y--S line and the E point $(k_x, \pi/b, \pi/c)$ on the T--R line belong to complex or pseudo-real representations, respectively, leading to degeneracy protected by time-reversal symmetry.
These degeneracies on the zone boundary correspond to the band-sticking effect mentioned above.
Along these lines, the two degenerate states at a given $\bm{k}$ on a screw axis (for example, along the $k_x$ direction) share the same sign and magnitude of the spin component along the axis ($s_x$), while their spin components perpendicular to the axis ($s_y$ and $s_z$) have equal magnitudes but opposite signs.
As a result, the averaged spin polarization of the two states is parallel to the screw axis.

Finally, we note a possible extension to chiral HOIP antiferromagnets in which global time-reversal symmetry is broken.
In such systems, the magnetic point group is reduced from $2221'$ to $222$ by magnetic ordering, allowing altermagnetism with no net magnetization ~\cite{Cheong2025} to emerge.
The above analysis suggests that exotic spin-split band structures may be realized in such antiferromagnets, where the degeneracies at the zone-boundary high-symmetry points and along the zone-boundary high-symmetry lines are lifted.

\section{\label{sec:summary}Summary}

In summary, we have performed comprehensive first-principles band-structure calculations for the chiral compounds (\textit{R}/\textit{S}-PEA)PbI$_3$ and (\textit{R}/\textit{S}-NEA)PbI$_3$, focusing on spin splitting induced by structural chirality and SOC.
We demonstrated that (\textit{R}/\textit{S}-NEA)PbI$_3$ exhibits stronger spin-splitting effects in the edges of the valence bands than (\textit{R}/\textit{S}-PEA)PbI$_3$, despite their similar linear-in-$k$ coefficients near high-symmetry points.
This enhancement originates from larger SOC-induced level splitting at high-symmetry points and (anti)crossings owing to the multiband nature and nonsymmorphic symmetry of the $P2_12_12_1$ space group.
Furthermore, we provide an explanation for the previously reported opposite CD signs in the two compounds with the same molecular handedness, either \textit{R} or \textit{S}.
We find that the two compounds with the same molecular handedness exhibit opposite signs of spin texture and of the structural chirality of the [PbI$_6$]$^{4-}$ octahedra.
Because the CD sign is closely related to the sign of the spin texture, these results indicate a consistent relationship among the signs of the CD, the spin-split band structure, and the structural chirality.
Our group-theoretical analysis clarified how nonsymmorphic symmetry governs band degeneracies, band-sticking, and the disappearance of spin polarization at specific Brillouin-zone boundaries.
These results provide a microscopic understanding of giant chiroptical responses in chiral HOIPs and establish key design principles for enhancing spin-dependent and chiroptical functionalities, including possible extensions to magnetic and antiferromagnetic chiral HOIPs.

\begin{acknowledgments}
We thank Yuya Ominato, Makoto Naka, and Shuntaro Sumita for fruitful discussions.
This work was supported by Japan Science and Technology Agency (JST) CREST (Grant No. JPMJCR23A1), and Japan Society for the Promotion of Science (JSPS) KAKENHI (Grant Nos. 22H01184, 23K17659, and 25K00963).
\end{acknowledgments}

\section*{Data Availability}
The data that support the findings of this article are not publicly available. The data are available from the authors upon reasonable request. The crystallographic information files used in the calculations are provided as Supplemental Material.

\bibliography{references}

\end{document}